\documentclass[pra,twocolumn,amsmath,amssymb,superscriptaddress,eqsecnum]{revtex4-2}
\usepackage{graphicx,relsize,color,mathtools,bm,newtxtext,newtxmath,braket,rotating,dsfont,mathtools}
\usepackage{float}

\usepackage[hyphenbreaks]{breakurl}
\usepackage[colorlinks=true,linkcolor=blue,citecolor=blue,urlcolor =blue]{hyperref}
\usepackage{float}
\usepackage{flafter}
\usepackage{placeins}
\usepackage{soul}
\usepackage{cancel}
\usepackage{comment}
\usepackage{amsmath}

\newcommand{\bd}[1]{\hat{b}_{#1}^{\dagger}}  
\newcommand{\bnd}[1]{\hat{b}_{#1}^{\phantom{\dagger}}}  
\newcommand{\ad}{\hat{a}^{\dagger}}  
\newcommand{\an}{\hat{a}^{\phantom \dagger}}  
\newcommand{\rhoOp}{\hat{\varrho}}  
\newcommand{\Prob}{P}  
\newcommand{\Wfunc}{W}  
\newcommand{\delp}[2]{\frac{\partial #1}{\partial #2}}  
\newcommand{\astconj}[1]{#1^{\ast}}  
\newcommand{\bn}[1]{\beta_{#1}}  
\newcommand{\al}{\alpha}  

\begin{document}

\title{Benchmarking quantum phase-space methods for near-resonant  light propagation}

\author{Mojdeh S. Najafabadi}  
\email{mojdeh.shikhali-najafabadi@mpl.mpg.de}
\affiliation{Max-Planck-Institut f\"{u}r die Physik des Lichts, 91058~Erlangen, Germany}       

\author{Joel F. Corney}
\affiliation{School of Mathematics and Physics, The University of Queensland, Brisbane, Queensland 4072, Australia} 
\author{Luis L. S\'anchez-Soto}
\affiliation{Max-Planck-Institut f\"{u}r die Physik des Lichts, 91058~Erlangen, Germany}       
\affiliation{Departamento de \'Optica, Facultad de F\'{\i}sica, Universidad Complutense, 28040 Madrid, Spain}
 \affiliation{Institute for Quantum Studies, Chapman University, Orange, CA~92866, USA}
        
\author{Gerd Leuchs}      
\affiliation{Max-Planck-Institut f\"{u}r die Physik des Lichts, 91058~Erlangen, Germany}        
\affiliation{Institut f\"{u}r Optik, Information und Photonik,  Friedrich-Alexander-Universit\"{a}t Erlangen-N\"{u}rnberg, 91058~Erlangen, Germany}

\begin{abstract}
We study the dynamics of light interacting with a near-resonant atomic medium using the truncated Wigner and positive $P$ phase-space representations. {By mapping the atomic degrees of freedom via the Jordan–Schwinger transformation, we first analyze the system under} unitary evolution {before extending the treatment to include} an optical reservoir. While both {methods} capture the main features of the light–matter dynamics, {our results demonstrate that the truncated Wigner approximation deviates significantly at higher interaction strengths and in regimes where reservoir-induced noise is prominent.}  
\end{abstract}

\maketitle{}

\section{Introduction} 

The phase-space formulation of quantum mechanics~\cite{Tatarskii:1983aa,Hillery:1984aa,Balazs:1984aa,Lee:1995aa,Schroek:1996aa,Ozorio:1998aa,Schleich:2001aa,QMPS:2005aa,Weinbub:2018aa,Rundle:2021aa} provides a conceptually transparent and self-contained alternative to the conventional Hilbert-space description. In this approach, quantum states are represented by quasiprobability distributions defined over a classical phase space, observables are mapped to $c$-number functions, and quantum dynamics is reformulated as stochastic evolution. This framework offers not only an intuitive visualization of quantum states and processes, but also practical computational tools for regimes where the direct solution of Heisenberg equations of motion or master equations becomes prohibitively complicated~\cite{Kampen:2011aa,Gardiner:2004aa,Breuer:2007aa}.

 For bosonic systems, several phase-space representations are widely employed, most notably the Wigner ($W$)~\cite{Wigner:1932aa}, Glauber–Sudarshan ($P$)~\cite{Glauber:1963aa,Sudarshan:1963aa}, and Husimi ($Q$)~\cite{Husimi:1940aa,Kano:1965aa} distributions. Despite their broad applicability, each of these representations suffers from significant limitations~\cite{Braunstein:1991aa}. In particular, some distributions become highly singular for physically relevant quantum states; for instance, the Glauber–Sudarshan $P$ function is more singular than a delta function for squeezed states and number eigenstates. Other representations, such as the Wigner function, may  take negative values, precluding its interpretation as a genuine probability density. Furthermore,  the equations governing their time evolution are not always true Fokker–Planck equations~\cite{Haken:1996aa}; they may involve derivatives of order higher than two or, even when restricted to second order, fail to possess a positive-definite diffusion matrix.
 
 To overcome these difficulties and enhance numerical applicability, particularly for complex and dissipative systems, several modified phase-space techniques have been developed~\cite{Walls:2008aa,Gardiner:2009aa}. Two of the most prominent methods are the positive $P$ representation (PPR) and the truncated Wigner approximation (TWA), each addressing different aspects of the challenges outlined above. 
 
The PPR, introduced by Drummond and Gardiner~\cite{Drummond:1980aa}, overcomes several limitations of earlier phase-space methods.  When the resulting evolution equation is of second order, the diffusion matrix can always be chosen as positive definite, yielding a \emph{bona fide} Fokker–Planck equation. Consequently, the probability distribution remains positive at all times, provided it is initialized with a suitable positive distribution, a condition that can always be satisfied~\cite{Gilchrist:1997aa,Schack:1991aa}. However, these benefits come at the cost of doubling the number of phase-space variables. More critically, this enlarged phase space can trigger numerical instabilities: stochastic noise may drive trajectories away from the stable ``classical" manifold into unstable regions where nonlinear drift terms amplify fluctuations. This can lead to boundary term problems or divergent trajectories in finite time. Despite these challenges, the PPR remains a cornerstone for simulating quantum effects in optics~\cite{Corney:2008aa} and ultracold atoms~\cite{Deuar:2007aa}.

The TWA~\cite{Blakie:2008aa,Polkovnikov:2010aa} provides a complementary approach, particularly suited to semiclassical regimes, where quantum fluctuations are small. To leading order, quantum effects enter only through the Wigner distribution of the initial conditions, while the subsequent dynamics is governed by classical equations of motion.  While computationally efficient and widely used for large-scale atomic systems~\cite{Steel:1998aa,Sinatra:2002aa,Dagvadorj:2015aa,Comaron:2018aa}, its accuracy relies on a systematic truncation that may fail in regimes of strong interactions, long evolution times, or low particle numbers.

Broadly speaking, the regimes of validity of the  PPR  and the  TWA are largely complementary: the TWA is typically accurate in the presence of strong  driving, whereas the PPR remains stable under strong dissipation~\cite{Hosseinabadi:2025aa}. This has motivated numerous comparative studies of the two techniques~\cite{Drummond:1993aa,Plimak:2001aa,Hoffmann:2008aa,Olsen:2009aa,Corney:2015aa,Ng:2011aa,Huber:2021aa}. Nevertheless, a comprehensive and systematic benchmark of these methods in open, strongly interacting light–matter systems (where intrinsic quantum fluctuations and reservoir-induced noise are  essential) has not yet been performed. Addressing this gap constitutes the central objective of this paper.

To this end, we consider a quantum model describing the propagation of radiation in an atomic  two-level medium subject to  radiative damping~\cite{Drummond:1991aa}. Using the PPR, we have previously applied this framework to the study of self-induced transparency (SIT)~\cite{Slusher:1974aa} in gas-filled hollow-core photonic crystal fibers, demonstrating the generation of amplitude~\cite{Najafabadi:2024aa} and quadrature squeezing~\cite{Najafabadi:2025aa}. While these results underscore the ability of the PPR to capture quantum optical effects in strongly nonlinear regimes, the underlying formulation relies on a collective description of large atomic ensembles to suppress higher-order terms arising from noncanonical commutation relations. This assumption restricts the generality of the model and entails a significant computational overhead. Moreover, it does not naturally accommodate the TWA, thereby limiting access to more efficient simulation strategies.

In this paper, we introduce an alternative formulation based on the Jordan–Schwinger mapping~\cite{Jordan:1935aa,Schwinger:1952aa}, which represents atomic operators in terms of bosonic modes. This transformation eliminates higher-order phase-space terms, enabling the direct application of the standard coherent-state PPR. Crucially, it also renders the TWA applicable to atomic systems, providing an efficient and scalable alternative to the more computationally demanding PPR~\cite{JNajafabadi:2025aa}.

Within this  framework, we derive two sets of stochastic differential equations—one for each representation—to describe coherent atom–field dynamics and the influence of optical noise. By comparing these simulations, we identify the specific parameter regimes where the TWA provides an accurate, low-cost alternative to the more demanding PPR.

\section{Model}
\label{sec:Rec}

We consider a model consisting of an ensemble of two-level atoms interacting with a single-mode propagating field (for example, in a waveguide). The $\mu$th atom is located at position $\mathbf{r}_{\mu}$ and has transition frequency $\omega_{\mu}$. To  model the medium, we discretize both space and frequency into lattice cells, {following the construction introduced in Ref.~\cite{Drummond:1991aa}}. Each cell is labeled by a composite index $n=(j,\nu)$, where $j$ denotes the spatial site and $\nu$ labels the atomic frequency. All atoms within a given spatial cell $j$ share the same position but may differ in resonance frequency due to inhomogeneous broadening. 

The set of atoms belonging to a given cell ${n}$ is denoted by $s({n})$. Each spatial cell has volume $\Delta V$ and the total number of atoms at this spatial cell is $N_j = \sum_\nu N_{{n}}$, obtained by summing over all frequency classes. For each spatio-frequency cell, we introduce collective atomic operators 
\begin{equation}
 \hat{J}^{+}_{{n}} = \sum_{\mu \in s(n)} \hat{\sigma}_{\mu}^{+} e^{i \mathbf{k}_0 \cdot \mathbf{r}_{\mu}} = ( \hat{J}^{-}_{{n}} )^{\dagger}    \, , 
   	\quad  \hat{J}^{z}_{{n}}  = \frac{1}{2} \sum_{\mu \in s(n)} \hat{\sigma}_{\mu}^{z} \, ,  
\end{equation}
where $\hat{\sigma}^{\pm}_{\mu}$ and $\hat{\sigma}^{z}_{\mu}$ are the Pauli operators for atom $\mu$, and $\mathbf{k}_0$ is the wave vector corresponding to the single-mode carrier frequency $\omega_0$. The phase factor accounts for the spatial dependence of the optical mode and will be absorbed into the definition of the collective operators hereafter. 

Within the rotating-wave and dipole approximations, the system Hamiltonian takes the form  $\hat{H} = \hat{H}_{\mathrm{A}} + \hat{H}_{\mathrm{F}} + \hat{H}_{\mathrm{AF}}$  with
\begin{equation}
    \begin{aligned}
        \hat{H}_{\mathrm{A}} & =  \hbar \sum_{n} \Delta_{\nu} \, \hat J_{{n}}^{z}  \, ,	  \\ 
	\hat{H}_{\mathrm{F}} & =  	\hbar \sum_j \sum_{j^{\prime}}\Delta \omega(j, j^{\prime}) \, \hat a^{\dagger}_{j} \hat{a}_{j^{\prime}}^{\phantom \dagger} \\
	\hat{H}_{\mathrm{AF}} & = \hbar g \sum_j \hat a_{j}^{\dagger}\sum_{\nu}\hat J^{-}_{{n}} +\mathrm{h.  c.} \, ,  
    \end{aligned}
\end{equation}
with h.c. denoting the Hermitian conjugate. The piece $\hat{H}_{\mathrm{A}}$ describes the atoms and $\Delta_{n} = \omega_{\nu} - \omega_{0}$ is the detuning of atoms in spatio-frequency cell $n$ from the optical carrier frequency $\omega_{0}$.  The field Hamiltonian $\hat{H}_{\mathrm{F}}$ describes propagation and dispersion of the guided mode, encoded in the coupling function  $\Delta \omega(j,j^{\prime}) = \omega(j,j^{\prime}) - \omega_{0}\,\delta^{(3)}_{j j^{\prime}}$, which couples different spatial sites.  The interaction Hamiltonian $\hat{H}_{\mathrm{AF}}$ describes the coupling between the guided field and the collective atomic excitations, with effective coupling strength that we assume to be identical for all the atoms and independent of the frequency and the wave vector.  For an ideal two-level system  this coupling reads~\cite{Allen:1975aa}
\begin{equation}
g = \left ( \frac{\omega_{0} | \mathbf{d} \cdot \mathbf{e}|^{2}}{2\hbar\varepsilon_0  V} \right )^{1/2} \, , 
\end{equation}
where $V$ is  the quantization mode,  $\mathbf{d}$ is the relevant dipole matrix element and $\mathbf{e}$ is the mode polarization.

Following the Jordan-Schwinger map~\cite{Jordan:1935aa,Schwinger:1952aa},  each two-level atom is represented using two independent bosonic modes, $\hat{b}_1$ and $\hat{b}_2$, corresponding to the ground and excited states, respectively. The Pauli operators are mapped onto these bosonic modes according to  
\begin{equation}
\hat{\sigma}^{+} = \bd{2} \bnd{1} , \quad 		
		\hat{\sigma}^{-} = \bd{1} \bnd{2},\quad 
		\hat{\sigma}^{z} = \frac{1}{2} (\bd{1} \bnd{1} - \bd{2} \bnd{2} ).
\end{equation}
The canonical commutation relations $[\bnd{i} , \bd{j}] = \delta_{ij}$ guarantee that these operators satisfy the standard angular-momentum algebra. 
In this representation, an atomic excitation corresponds to annihilating a ground-state boson and creating an excited-state boson. 

Extending this mapping to the collective operators, we obtain 
\begin{equation}
	\hat{J}^{+}_{{n}} = \bd{2n} \bnd{1n}  =	(\hat{J}^{-}_{{n}} )^{\dagger}  \, ,
	\qquad
	\hat{J}^{z}_{{n}} = \frac{1}{2} (\bd{1n} \bnd{1n} - \bd{2n} \bnd{2n} )\, ,
\end{equation} 
where, for example, $\bd{1n}\bnd{1n} = \sum_{\mu \in s(n)} \bd{1\mu} \bnd{1\mu}$ counts the number of ground-state atoms within  cell $n$.
{

Strictly speaking, the Jordan--Schwinger representation is equivalent
to the finite-dimensional atomic Hilbert space only within the subspace
satisfying the local boson-number constraint
\begin{equation}
\bd{1n}\bnd{1n}+\bd{2n}\bnd{2n}=N_n ,
\end{equation}
which fixes the total number of bosons to the number of atoms in the cell.

In the simulations presented below, the coherent-state initial conditions
do not strictly satisfy this constraint at the level of individual stochastic
realizations, since coherent states are superpositions of different boson-number
sectors. However, the dynamics conserve the total particle number, such that
different number sectors evolve independently without generating coherences
between them. The resulting fluctuations therefore remain small in the
large $N_n$ regime considered here and do not significantly affect the
dynamics or the conclusions of this work.
}

In terms of these bosonic operators, the  interaction Hamiltonian  becomes
	\begin{equation}
		\hat{H}_{\mathrm{int}} =   \hbar {g}  \sum_{j,\nu}  ( {\hat a_{j}^{\dagger}}   \bd{1n} \bnd{2n} 
		+   \bd{2n} \bnd{1n} {\hat a_j}) \,. 
	\end{equation}
 
 \begin{table}[t]
 	\centering
 	\renewcommand{\arraystretch}{1.5}  
 	\caption{Mapping Rules for PPR and TWA}
 	\label{table:mapping_rules}
 	\begin{tabular}{lcc}
 		\hline
 		\textbf{Operator} & \textbf{PPR} & \textbf{TWA} \\
 		\hline
 		$\an \rhoOp$ & $\alpha P$ & $\alpha + \frac{1}{2} \frac{\partial}{\partial \alpha^\ast} W$ \\
 		$\ad \rhoOp$ & $(\alpha^{+}-\frac{\partial}{\partial \alpha}) P$ & $\alpha^{\ast} - \frac{1}{2} \frac{\partial}{\partial \alpha} W$ \\
 		$\rhoOp \ad$ & $\alpha^{+} P$ & $\alpha^{\ast}+\frac{1}{2} \frac{\partial}{\partial \alpha} W$ \\
 		$\rhoOp \an$ & $(\alpha - \frac{\partial}{\partial \alpha^{+}}) P$ & $\alpha - \frac{1}{2} \frac{\partial}{\partial \alpha^*} W$ \\
 		\hline
 	\end{tabular}
 \end{table}
  
   In addition to the interaction with resonant atoms, the optical mode may also interact with nonresonant atoms within the medium, leading to absorption.  We model this absorption with nonresontant dipoles in which saturation and coherent effects can be neglected.  By tracing out the absorbing dipoles,  the resulting dynamics of the density operator $\rhoOp$ is governed by the master equation
\begin{align}
    \frac{\partial \rhoOp}{\partial t} 
    = \frac{1}{i \hbar} [ \hat{H} , \rhoOp ] + L_{\text{opt}} [ \rhoOp ] \, ,
\end{align}
with the Lindblad superoperator $L_{\text{opt}}$ is~\cite{Breuer:2007aa}
\begin{align}
    L_{\text{opt}} [\rhoOp]  
    &= \frac{\gamma}{2} \sum_{j}   [  (1+\Bar{n}) 
     ( [ \an_{j} \rhoOp, \ad_{j}] + [\an_{j}, \rhoOp \ad_{j}]  ) \notag \\
    & +  \Bar{n}  ([\ad_{j} \rhoOp, \an_{j}] + [\ad_{j}, \rhoOp \an_{j}] )  ] \, ,
\end{align}  
where $\gamma$ represents the optical loss rate and {$\bar{n}$ denotes the effective thermal occupation number associated with the non-resonant atomic reservoir.}

While not explicitly treated in this paper, this model is robust enough to incorporate additional atomic dissipation processes. Spontaneous emission into nonguided channels or dephasing effects can be included by integrating the appropriate radiative and collisional reservoirs into the master equation framework~\cite{Drummond:1991aa}.

To analyze the system dynamics, we map the density operator onto a phase-space distribution ($P$ or $W$) using the standard correspondence rules for bosonic operators detailed in Table~\ref{table:mapping_rules}. This procedure yields a Fokker–Planck equation of the general form~\cite{Gardiner:2004aa}:
\begin{align}
	\label{Eq:FPE}
    \frac{\partial P(\bm{\alpha})}{\partial t} 
    & = -\sum_q \frac{\partial}{\partial \alpha_q}\!\big[A_q(\bm{\alpha})P(\bm{\alpha})\big] \nonumber\\
      &+ \frac{1}{2}\sum_{qq'}\frac{\partial^2}{\partial \alpha_q \partial \alpha_{q'}}\!\big[D_{qq'}(\bm{\alpha})P(\bm{\alpha})\big],
\end{align}
where $A_q(\bm{\alpha})$ are the drift coefficients, $D_{qq'}(\bm{\alpha})$ are the diffusion coefficients, and $\alpha_q$ are the phase-space variables, collectively denoted $\bm{\alpha}$.  

The second-order derivative terms in Eq.~\eqref{Eq:FPE} characterize the evolution as a diffusion process. Following the standard equivalence between Fokker–Planck equations and Itô stochastic differential equations~\cite{Gardiner:2004aa}, Eq.~\eqref{Eq:FPE} can be mapped onto a set of Langevin equations where the number of equations is proportional to the number of bosonic modes:
\begin{equation}
    d\alpha_{q} = A_{q}(\bm{\alpha})\,dt + \sum_{q'} B_{qq'}(\bm{\alpha})\, d\eta_{q'}(t) \, .
\end{equation}
In this representation, the diffusion matrix $\mathbf{D}$ is factorized as $\mathbf{D} = \mathbf{B} \mathbf{B}^{\top}$, $\top$ indicating the transpose. This decomposition is not unique, allowing for a  freedom in the choice of the noise matrix $\mathbf{B}$. In the  PPR, this flexibility enables different stochastic realizations of the same underlying quantum dynamics.

The stochastic driving terms, $d\eta$, are Wiener processes. These provide the appropriate representation of the noise, as they are characterized by Gaussian increments with a covariance proportional to $dt$. Specifically, they satisfy
\begin{align}
    &\langle d\eta_{q}(t) \rangle = 0, \notag \\
    &\langle d\eta_{q}(t)\, d\eta_{q'}(t^{\prime}) \rangle =  0,     \\
    &\langle d\eta_{q}(t)\, d\eta_{q'}(t) \rangle = \delta_{qq'}\,dt.   \notag
\end{align}

\section{Dynamics of the positive $P$ distribution}
\label{II}

In the PPR, the density operator is expanded in an overcomplete basis of coherent states.  This construction maps the operator equations of motion onto $c$-number stochastic differential equations defined on an enlarged phase space, in which each complex variable is paired with an independent conjugate variable. The resulting distribution function $\Prob$ is normalized over the full complex phase space and can be interpreted as a quasiprobability density $P(\bm{\alpha}, t)$.  Here $\bm{\alpha} = (\alpha_j,\alpha_j^{+},\beta_{1n},\beta_{1n}^{+},\beta_{2n},\beta_{2n}^{+})^{\top}$ denotes the full set of phase-space variables. This representation is particularly powerful as it allows nonclassical states and nonlinear atom–light interactions to be treated in a numerically tractable manner. However, as discussed below, these simulations can be susceptible to numerical instabilities at long integration times.

Using the operator correspondences listed in Table~\ref{table:mapping_rules}, we obtain a Fokker–Planck equation for the positive 
$P$ distribution containing first- and second-order derivatives,
  \begin{equation}
        \begin{aligned}
       \label{Eq:Diff_PPR} 
       \frac{\partial}{\partial t}\Prob(\bm{\alpha}, t)  &= \sum_{j,j'}\delp{}{\al_j} 
         i \Delta \omega(j,j^{\prime}) \al_{j^{\prime}} \Prob(\bm{\alpha}, t)    \\
        &+  \sum_{j,\nu} \delp{}{\al_j} 
        i{g}\left(  {\bn{1n}}^{+} \bn{2n} 
        +  \delp{}{\bn{1n}} \bn{2n}
        \right ) \Prob(\bm{\alpha}, t)   \\
       &+  \sum_{j,\nu}\delp{}{\bn{1n}} \left ( 
         i{g} {\al_j}^{+} \bn{2n} 
        + \frac{i}{2} \Delta_{\nu}  \bn{1n} 
        \right ) \Prob(\bm{\alpha}, t)  \\
        &  +  \sum_{j,\nu}\delp{}{\bn{2n}} \left ( 
         i{g} \al_j \bn{1n} 
        - \frac{i}{2} \Delta_{\nu}  \bn{2n}  \right ) \Prob(\bm{\alpha}, t)  \\
        &   + \sum_j\delp{}{\al_j} \gamma \left (
         \frac{1}{2} \al_j +
        \Bar{n} \delp{}{{\al}^{+}_{j}}  \right ) \Prob(\bm{\alpha}, t)  + {\rm c.c.}
    \end{aligned}
  \end{equation}

There are no mixed derivatives involving variables belonging to different spatial or frequency cells, reflecting the locality of both the atom–light interaction and the optical reservoir. The diffusion matrix therefore has a block-diagonal structure, with one block per cell 
$n=(j,\nu)$:
\begin{equation}
	\label{DiffMatrix_PPR}
    D_n =
    \begin{pmatrix}
        0  & D^{\alpha_j \alpha_j^{+}} & D^{\alpha_j \beta_{1n}} & 0 & 0  \quad & 0 \\[1pt]
        D^{\alpha_j^{+}\alpha_j} & 0 & 0 & D^{\alpha_j^{+}\beta_{1n}^{+}} & 0   \quad& 0 \\[1pt]
        D^{\beta_{1n}\alpha_j} & 0 & 0 & 0 & 0  \quad & 0 \\[1pt]
        0 & D^{\beta_{1n}^{+}\alpha_j^{+}} & 0 & 0 & 0 \quad & 0 \\[1pt]
        0   & 0 & 0 & 0 & 0 \quad & 0 \\
        0 & 0 & 0 & 0 & 0 \quad & 0    
    \end{pmatrix},
\end{equation}
which contains contributions from both the atom–light coupling and the optical reservoir. Accordingly, we write $\mathbf{D} = \mathbf{D}^{(0)}  + \mathbf{D}^{(1)}$, where $\mathbf{D}^{(0)}$   arises from the coherent atom–light interaction and  $\mathbf{D}^{(1)}$  from coupling to the reservoir. If each contribution can be factorized independently as $\mathbf{D}^{(i)} = \mathbf{B}^{(i)} \mathbf{B}^{(i) \top}$,  then the total noise matrix $\mathbf{B}$ can be written as  $\mathbf{B} = (\mathbf{B}^{(0)}  \; \mathbf{B}^{(1)} )$, which gives  a noise matrix for each cell of the form
\begin{widetext}
\begin{align}
\mathbf{B}_n =
    \begin{pmatrix}
        -\sqrt{D^{\alpha_j \alpha_j^{+}}/2}  & -i\sqrt{D^{\alpha_j \alpha_j^{+}}/2} 
        & -\sqrt{D^{\alpha_j \beta_{1n}}/2} & -i\sqrt{D^{\alpha_j \beta_{1n}}/2} 
        & 0 & 0 \\
        -\sqrt{D^{\alpha_j \alpha_j^{+}}/2}  & i\sqrt{D^{\alpha_j \alpha_j^{+}}/2} 
        & 0 & 0 
        & -\sqrt{D^{\alpha_j^{+}\beta_{1n}^{+}}/2} & -i\sqrt{D^{\alpha_j^{+}\beta_{1n}^{+}}/2} \\
        0 & 0 
        & -\sqrt{D^{\alpha_j \beta_{1n}}/2} & i\sqrt{D^{\alpha_j \beta_{1n}}/2} 
        & 0 & 0 \\
        0 & 0 
        & 0 & 0 
        & -\sqrt{D^{\alpha_j^{+}\beta_{1n}^{+}}/2} & i\sqrt{D^{\alpha_j^{+}\beta_{1n}^{+}}/2} \\
        0 & 0 & 0 & 0 & 0 & 0 \\
        0 & 0 & 0 & 0 & 0 & 0    
    \end{pmatrix} \, .
\end{align}
\end{widetext}

The atom–light coupling requires independent noise processes to drive variables that are nominally complex conjugates. This is accommodated in the PPR by doubling the phase space, such that the pairs $(\alpha_j , \alpha_j^{+})$, $(\beta_{1n}, \beta_{1n}^{+})$, and 
$(\beta_{2n},  \beta_{2n}^{+})$ are treated as independent variables rather than strict complex conjugates. We therefore use the superscript 
$+$ instead of the complex conjugation symbol $\ast$. Further details may be found in
Ref.~\cite{Gardiner:2004aa}.

The corresponding Itô stochastic differential equations are
\begin{equation}
\begin{aligned}
    \frac{d\alpha_j}{dt} &= -\frac{\gamma}{2}\,\alpha_j
    - i\,\sum_{j'}\Delta\omega(j,j^{\prime})\,\alpha_{j'}
    - i g \sum_{\nu} \beta_{1n}^{+}\beta_{2n} 
    + F^{\alpha_j},   \label{Eq:Alpha_Evolution}\\
    \frac{d\alpha_j^{+}}{dt} &= -\frac{\gamma}{2}\,\alpha_j^{+}
    + i\,\sum_{j'}\Delta\omega(j,j^{\prime})\,\alpha_{j'}^{+}
    + i g \sum_{\nu} \beta_{2n}^{+}\beta_{1n} 
    + F^{\alpha_j^{+}},  \\
    \frac{d\beta_{1n}}{dt} &=  
    - i g \alpha_j^{+} \beta_{2n} 
    - \frac{i}{2} \Delta_{\nu}\,\beta_{1n} 
    + F^{\beta_{1n}},  \\
    \frac{d\beta_{1n}^{+}}{dt} &=
    + i g \alpha_j \beta_{2n}^{+} 
    + \frac{i}{2} \Delta_{\nu}\,\beta_{1n}^{+} 
    + F^{\beta_{1n}^{+}},  \\
    \frac{d\beta_{2n}}{dt} &=  
    - i g \alpha_j \beta_{1n} 
    + \frac{i}{2}\Delta_{\nu}\,\beta_{2n} 
    + F^{\beta_{2n}},  \\
    \frac{d\beta_{2n}^{+}}{dt} &= 
    + i g \alpha_j^{+} \beta_{1n}^{+} 
    -  \frac{i}{2}\Delta_{\nu}\,\beta_{2n}^{+} 
    + F^{\beta_{2n}^{+}}. 
\end{aligned}
\end{equation}
Quantum  fluctuations are entirely captured by the stochastic terms $F$, which are influenced
 by the properties of the reservoirs and the nonlinear coupling between the atoms and the radiation field. 
 These terms can be expressed using fundamental Gaussian stochastic processes as follows:
 \begin{equation}
 \begin{aligned}
 	F^{\alpha_j} &= \sqrt{\frac{\gamma \bar{n}}{2}} \, (d\eta^{1}_{j} + i d\eta^{2}_{j} )
 	+ \sum_\nu \sqrt{\frac{-i g\beta_{2n}}{2}} \, (d\eta_{n}^{3} + i d\eta_{n}^{4} ),  \\
 	F^{\alpha_j^{+}} &= \sqrt{\frac{\gamma \bar{n}}{2}} \, (d\eta^{1}_{j} - i d\eta^{2}_{j})
 	+  \sum_\nu \sqrt{\frac{i g\beta_{2n}^{+}}{2}} \,(d\eta_{n}^{5} + i d\eta_{n}^{6}),  \\
 	F^{\beta_{1n}} &= \sqrt{\frac{-i g\beta_{2n}}{2}} \, (d\eta_{n}^{3} - i d\eta_{n}^{4}),  \\
 	F^{\beta_{1n}^{+}} &= \sqrt{\frac{i g\beta_{2n}^{+}}{2}} \, (d\eta_{n}^{5} - i d\eta_{n}^{6}),  \\
 	F^{\beta_{2n}} &= F^{\beta_{2n}^{+}} = 0,  
 \end{aligned}   
 \end{equation}
 where $\eta_j^1$, $\eta_j^2$,  $\eta_n^3,  \dots , \eta_n^6$ are real Wiener increments.    

\section{Dynamics in the truncated Wigner approximation}
\label{III}

In this section, we reformulate the master equation for the density matrix $\rhoOp$ in terms of the Wigner  distribution $\Wfunc(\bm{\alpha}, t)$.  We apply again the rules in Table~\ref{table:mapping_rules} and retain terms up to second-order derivatives, which gives us the Fokker-Planck equation 
\begin{equation}
    \begin{aligned}
    	\label{Eq:FPE_Wigner}
         \delp{}{t}\Wfunc(\bm{\al}, t)  & = \sum_{j,j'} \delp{}{\al_j} 
        i  \Delta \omega(j,j^{\prime}) \al_{j^{\prime}}    \Wfunc(\bm{\al}, t)  \\
        & + \sum_{j,\nu} \delp{}{\al_j}  i g \astconj{\bn{1n}}\bn{2n} 
        \Wfunc(\bm{\al}, t)   \\
       & +  \sum_{j, \nu}\delp{}{\bn{1n}} \left (  
        ig \astconj{\al_j} \bn{2n} 
       +\frac{i}{2} \Delta_{\nu}  \bn{1n} 
        \right ) \Wfunc(\bm{\al}, t)  \\
        & +  \sum_{j, \nu} \delp{}{\bn{2n}} \left (
           ig  \al_j \bn{1n}
        -\frac{i}{2}  \Delta_{\nu} \bn{2n} 
        \right )\Wfunc(\bm{\al}, t)  \\
      & + \sum_{j}\delp{}{\al_j} \gamma \left ( 
          \frac{1}{2}\al_j  + \left(\Bar{n}+1/2 \right)\delp{}{\astconj{\al_j}}
        \right ) \Wfunc(\bm{\al}, t) + {\rm c.c.}
    \end{aligned}
   \end{equation}

In contrast with the PPR, the Wigner representation produces derivative terms of order higher than two stemming from the interaction Hamiltonian. While the second-order interaction noise terms cancel identically, the remaining third-order derivatives are neglected in the  TWA. This procedure is consistent with the general structure of the Wigner equation, in which purely Hamiltonian evolution generates only odd-order derivatives~\cite{Corney:2015aa}.

Importantly, neglecting these higher-order terms does not mean that interaction-induced quantum fluctuations are lost. Rather, such fluctuations are incorporated through the stochastic sampling of the initial Wigner distribution, which contains the appropriate vacuum and thermal noise. As the system evolves, the nonlinear dynamics transform this initial uncertainty, enabling the TWA to capture leading-order quantum effects even in the absence of explicit interaction noise terms in the equations of motion. 

The Itô equations are:
\begin{equation}
\begin{aligned}
    \frac{d\alpha_j}{dt} &=
        - \frac{\gamma}{2}\,\alpha_j 
        - i\sum_{j'}\Delta\omega(j,j^{\prime})\,\alpha_j
        - i g \sum_{\nu} \beta_{1n}^{\ast}\beta_{2n} 
        + F^{\alpha_j},  \\
    \frac{d\alpha_j^{\ast}}{dt} &=
        -  \frac{\gamma}{2}\,\alpha_j^{\ast}
        + i\sum_{j'}\Delta\omega(j,j^{\prime})\,\alpha_j^{\ast}
        + i g \sum_{\nu} \beta_{1n}\beta_{2n}^{\ast} 
        + F^{\alpha_j^{\ast}},  \\
    \frac{d\beta_{1n}}{dt} &=
        - \tfrac{i}{2}\Delta_{\nu}\,\beta_{1n} 
        - i g\,\alpha_j^{\ast}\beta_{2n},  \\
    \frac{d\beta_{1n}^{\ast}}{dt} &=
         \tfrac{i}{2}\Delta_{\nu}\,\beta_{1n}^{\ast}
        + i g\,\alpha_j \beta_{2n}^{\ast},  \\
    \frac{d\beta_{2n}}{dt} &=
         \tfrac{i}{2}\Delta_{\nu}\,\beta_{2n}
        - i g\,\beta_{1n}\alpha_j,  \\
    \frac{d\beta_{2n}^{\ast}}{dt} &=
        - \tfrac{i}{2}\Delta_{\nu}\,\beta_{2n}^{\ast}
        + i g\,\beta_{1n}^{\ast}\alpha_j^{\ast} \, ,
\end{aligned}
\end{equation}
with optical noise defined as
\begin{equation}
   F^{{\al}_j} =  -\sqrt{\tfrac{1}{2}  \gamma ( \bar{n} +1/2 )} \;  (d\eta_j^0 +id\eta_j^1) = ({F^{{\al}_j}}^{\ast})^{\ast} \, .
\end{equation}

\section{Transition to continuous variables}

We now pass to the continuum limit by associating the discrete optical amplitudes with a continuous field,
\begin{equation}
\alpha_j(t) \mapsto \alpha(t,\mathbf{r}_j),
\qquad
\sum_j \to \int \frac{d^3\mathbf{r}}{\Delta V},
\end{equation}
and subsequently taking the cell volume  $\Delta V \to 0$.

The free propagation of the optical field is governed by the discrete coupling operator $\Delta\omega(j,j')$, which mixes field amplitudes at different lattice sites.  Under the slowly varying envelope (SVEA) and paraxial approximations--corresponding to a narrow optical bandwidth compared to the carrier frequency  $\omega_{0}$,  weak longitudinal variation, and small transverse propagation angles, the discrete sum becomes the differential operator ({ see appendix A})
\begin{equation}
  i \sum_{j'} \Delta \omega(j,j')\, \alpha_{j'}(t)
  \;\simeq\;
  v_g\,\frac{\partial}{\partial z}\,\alpha(t,\mathrm{x})
  - i\,\frac{v_g}{2k_0}\,\nabla_{\!\perp}^{2}\,\alpha(t,\mathrm{x}),
  \label{eq:svea_paraxial}
\end{equation}
where $z$ is the longitudinal propagation coordinate, $\nabla_{\perp}^{2}$ is the transverse Laplacian, and $v_g$ denotes the group velocity in the waveguide in the absence of the atomic medium.

For a single guided transverse mode, we can decompose the fields as $\alpha(t,\mathbf{r}) = \alpha(t,z)\,u(\mathbf{r}_\perp)$. By projecting onto the normalized mode profile $u(\mathbf{r}_\perp)$, the transverse dynamics is absorbed into an effective mode area.  In this single-mode approximation, the transverse Laplacian vanishes,  reducing the optical field to a one-dimensional propagation equation along the longitudinal axis $z$. 

While a fully three-dimensional treatment would account for atoms experiencing spatially varying intensities across the transverse profile, we here restrict the atomic degrees of freedom to a one-dimensional grid.  This simplification facilitates a direct comparison between the different phase-space methods considered above. Each grid cell is taken to have a longitudinal extent $\Delta L$ and an effective transverse area $A_{\rm eff}$.

For the description of propagating optical pulses, it is convenient to work in a co-moving reference frame. We therefore introduce the retarded time variable $\tau = t - z/v_g$, { which gives the timing relative to a pulse travelling at group velocity  $v_g$. } This transformation confines the dynamics to a narrow temporal window and substantially reduces the computational effort~\cite{Corney:2008aa}.  We also  introduce the photon-flux field
\begin{equation}
\phi(\tau,z) = \sqrt{\frac{v_g}{\Delta L}}\;\alpha(t,z),
\end{equation}
 which is normalized such that  $\phi^\dagger \phi$ gives the number of photons crossing the position $z$ per unit time. The one-dimensional atomic density $\rho_{1\mathrm{D}}$ is
\begin{equation}
    \rho_{1\mathrm{D}}(z,\omega )  = \int d^2\mathbf{r}_\perp\,\rho(\mathbf{r},\omega ) \, ,
\end{equation}
where $ \rho(\mathbf{r},\omega) = \frac{N_j}{\Delta V}f_{\omega}(\omega)$ is the density of resonant atoms in position $\mathbf{r}$ and a frequency $\omega$ with $f(\omega)$ to be the spectral lineshape of atoms.

In terms of $\phi$, the propagation equation~\eqref{Eq:Alpha_Evolution} and the atomic polarization field $R^\pm$ read as 

\begin{equation}
\begin{aligned}\label{Eq:phi}
 \frac{\partial}{\partial z}\phi(\tau,z) & =
 -\frac{\kappa}{2}\,\phi(\tau,z)  - i g_\phi \int d\omega\,\rho_{1\mathrm{D}}(z,\omega) \, R^{-}(\tau,z,\omega)   \\
 & + F^{\phi}(\tau,z) \, ,\\
  \frac{\partial}{\partial \tau} R^{-}(\tau,z,\omega)
&=
i \Delta_{\omega}\, R^{-}(\tau,z,\omega)
+ 2 i g_\phi \, \phi(\tau,z)\, R^{z}(\tau,z,\omega)\\
&+ F^{R^{-}}(\tau,z,\omega),
\\
\frac{\partial}{\partial \tau} R^{+}(\tau,z,\omega)
&=
- i \Delta_{\omega}\, R^{+}(\tau,z,\omega)
- 2 i g_\phi \, \phi^{+}(\tau,z)\, R^{z}(\tau,z,\omega) \\
&+ F^{R^{+}}(\tau,z,\omega).
\\
 \frac{\partial}{\partial \tau} R^z(\tau,z,\omega)
&=
i g_\phi\left[
\phi^+(\tau,z)\, R^-(\tau,z,\omega)
-
\phi(\tau,z)\, R^+(\tau,z,\omega)
\right] \\
&+ F^{R^z}(\tau,z,\omega)
\end{aligned}
\end{equation}
 where the effective one-dimensional atom–field coupling constant is
\begin{equation}
g_\phi = \frac{g}{2}\sqrt{\frac{\Delta L}{v_g}} = \frac{d}{2}\sqrt{\frac{\omega_0}{2\hbar \varepsilon_0  v_g A_{\rm eff}}}.
\end{equation}
Here the atomic polarization fields $R^\pm$ and population field $R^z$ are defined as
\begin{equation}
\begin{aligned}\label{Eq:R_min}
R^{-}(\tau,z,\omega)  & = \frac{2}{N_n}\,\beta_{1n}^+(\tau)\beta_{2n}(\tau), \\
R^{+}(\tau,z,\omega)  &= \frac{2}{N_n}\,\beta_{2n}^+(\tau)\beta_{1n}(\tau),\\
R^z(\tau,z,\omega)
&= \frac{1}{N_n}
\left[
\beta_{2n}^+(\tau)\beta_{2n}(\tau)
-
\beta_{1n}^+(\tau)\beta_{1n}(\tau)
\right].
\end{aligned}
\end{equation}

In the continuum limit, the fluctuations become delta correlated in time and space, with, for example, the optical reservoir  noise associated with the photon-flux field satisfying
\begin{equation}
\langle F^{\phi}(\tau,z)\,F^{\phi\dagger}(\tau^\prime,z^\prime) \rangle = \kappa\left(\overline{n} + s\right)
\,\delta(\tau-\tau')\delta(z-z'),
\end{equation}
where $\kappa=\gamma/v_g$ is the optical attenuation coefficient. The parameter $s$ accounts for the ordering of the phase-space representation:  $s=0$ for the normally ordered PPR and $s = 1/2$ for the symmetrically ordered TWA.
{

In the PPR, the nonlinear atom--field coupling additionally generates cross-correlations between the optical and atomic noise terms, arising from shared Wiener increments. Using Eq.~\eqref{Eq:R_min}, one finds
\begin{widetext}
\begin{equation}
    \begin{aligned}\label{Eq:Atom_field_Int}
\langle F^{\phi}(\tau,z)\,F^{R^{+}}(\tau',z',\omega)\rangle  & = -\,i\,g_{\phi}\,
[ 1 +  \langle R^{z}(\tau,z,\omega) \rangle ] \delta(\tau-\tau')\delta(z-z'), \\
\langle F^{\phi\dagger}(\tau,z)\,F^{R^{-}}(\tau',z',\omega)\rangle  & = \,i\,{g}_{\phi}\,
(1+\langle R^{z}(\tau,z,\omega)\rangle)\,
\delta(\tau-\tau')\delta(z-z'), \\
\langle F^{\phi}(\tau,z)\,F^{R^{z}}(\tau',z',\omega)\rangle &=  i g_{\phi} \langle R^{-}(\tau,z,\omega) \rangle \delta(\tau-\tau')\delta(z-z'), \\
\langle F^{\phi^{\dagger}}(\tau,z)\,F^{R^{z}}(\tau',z',\omega)\rangle &= -i g_{\phi} \langle R^{+}(\tau,z,\omega) \rangle \delta(\tau-\tau')\delta(z-z').
\end{aligned}
\end{equation}
\end{widetext}
It is these stochastic terms that, for example, enable spontaneous emission of the atoms into the guided mode.  In the TWA, where such noise terms are absent, spontaneous emission effects arise due to the inclusion of vacuum noise in the initial conditions (see below).

To complete the specification of the PPR noise terms, we note that the noises entering the various atomic variables are uncorrelated with each other:
\begin{widetext}
\begin{equation}
 \begin{aligned}
\langle F^{R^{+}}(\tau,z,\omega)\, F^{R^{-}}(\tau',z',\omega') \rangle &=  \langle F^{R^{z}}(\tau,z,\omega)\, F^{R^{z}}(\tau',z',\omega') \rangle = 0, \\ 
\langle F^{R^{z}}(\tau,z,\omega)\, F^{R^{\pm}}(\tau',z',\omega') \rangle &=  \langle F^{R^{z}}(\tau,z,\omega)\, F^{\phi}(\tau',z') \rangle = 0.
\end{aligned}
\end{equation}
\end{widetext}}
\section{Benchmarking phase-space methods}
\begin{figure}[t]
	\centering\includegraphics[width=\columnwidth]{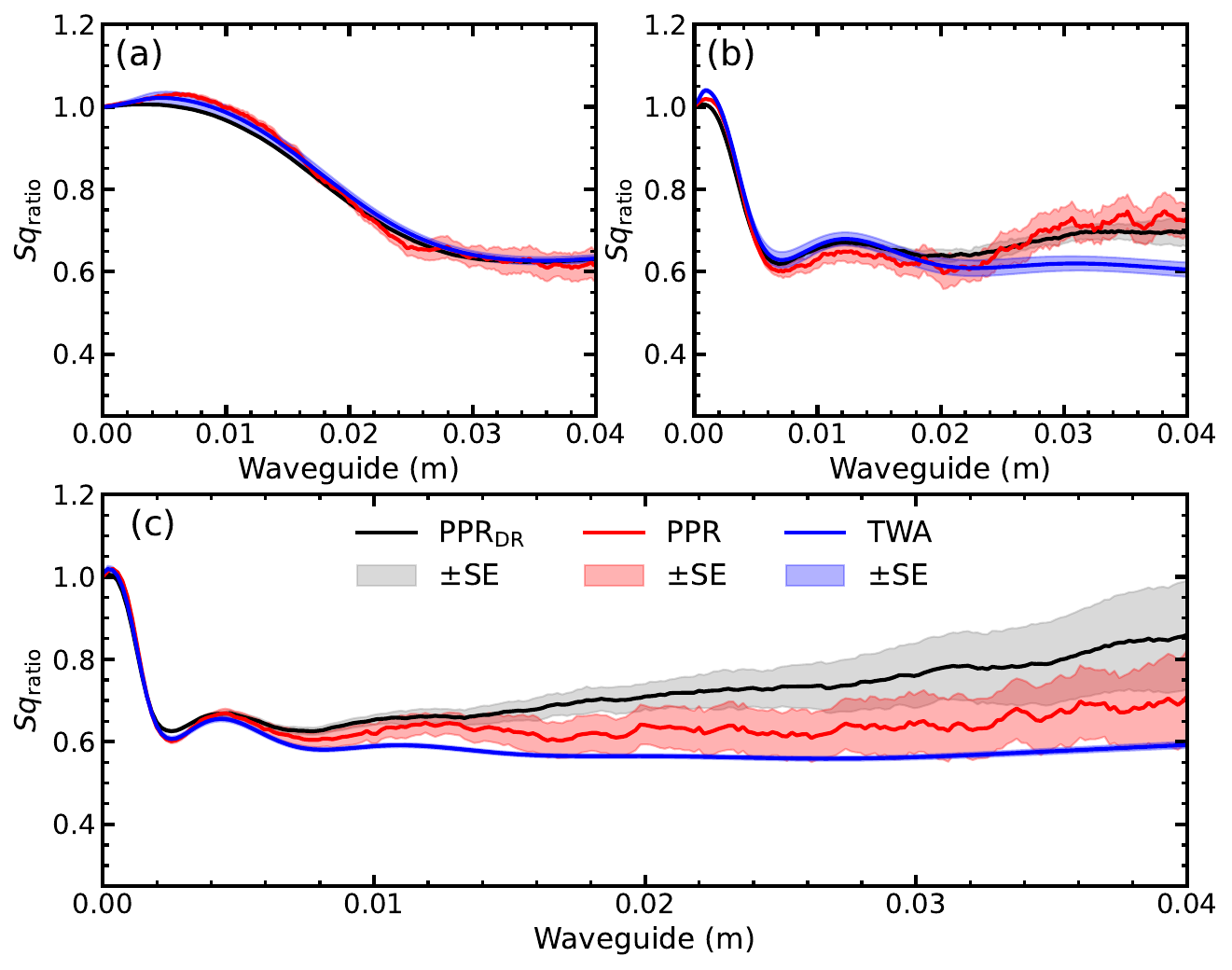}
	\caption{Calculation of the squeezing ratio for the optimum angle of local oscillator in the absence of an optical reservoir as a function of propagation distance using three different methods: PPR (red), TWA (blue), and $\mathrm{PPR}_{\mathrm{D}}$ (black). Shaded regions indicate sampling error. The total atomic densities are (a) $\rho = 2.65 \times10^{21}\,\mathrm{m^{-3}}$, (b) $\rho = 1.33\times10^{22}\,\mathrm{m^{-3}}$, and
		(c) $\rho = 3.7\times10^{22}\,\mathrm{m^{-3}}$. The number of stochastic trajectories used is approximately $2\times10^{5}$ for PPR, {$3\times10^{4}$–-$6\times10^{4}$ for $\mathrm{PPR}_{\mathrm{DR}}$}, and $5\times10^{3}$ for TWA.}
	\label{fig:Fig_all_methods_Unitray}
\end{figure}

In this section, we benchmark the equations derived within the TWA against the Itô equations obtained from the  PPR { i.e., Eq.~\eqref{Eq:phi} to Eq.~\eqref{Eq:Atom_field_Int}}.  The comparison is carried out under identical physical conditions in order to delineate the regimes of validity and the limitations of each method.  

To initialize the simulations in the retarded-time framework introduced above, it is necessary to specify both the optical input at the entrance of the waveguide and the initial atomic state throughout the medium. For the PPR, we consider initial conditions corresponding to a $2\pi$ SIT soliton, given by 
\begin{equation}
\begin{aligned}
R^{-}(0)&= R^+(0) = 0 \\
R^z(0) &= -1,\\
\phi(\tau,0) &=
\frac{A}{g_\phi}\,\mathrm{sech}\!\left[A(\tau-\tau_0)\right] ,
\label{Eq:coherentPulse}
\end{aligned}
\end{equation}
where $1/A$ sets the pulse width and $\tau_0$ is a timing offset (which we set to zero in the simulations below). 

For the TWA, quantum fluctuations are incorporated through stochastic initial conditions that sample the Wigner distribution,

\begin{equation}
\begin{aligned}
R^{-}(0) &= \frac{2}{\sqrt{N_n}}\delta\beta_{2n} + \frac{2}{N_n}\delta \beta^{\ast}_{1n} \delta \beta_{2n}, \\
R^{+}(0) &= 
\frac{2}{\sqrt{N_n}}\delta\beta_{2n}^{\ast}
+ \frac{2}{N_n}\delta\beta_{2n}^{\ast}\delta\beta_{1n},\\
\phi(\tau,0) &=
\frac{A}{g_\phi}\,\mathrm{sech}\!\left[A(\tau-\tau_0)\right]
+ \delta\phi(\tau),\\
R^z(0) &= -1-\frac{1}{\sqrt{N_n}}(\delta \beta_{1n} + \delta \beta^{\ast}_{1n})
\end{aligned}
\end{equation}

with noise correlations

\begin{equation}
\begin{aligned}
\langle \delta\phi(\tau)\delta\phi^\ast(\tau^\prime)\rangle
&= \tfrac{1}{2}\delta(\tau-\tau^\prime),\\
\langle \delta\beta_{1n}\delta\beta_{1n'}^\ast \rangle
&= \langle \delta\beta_{2n}\delta\beta_{2n^\prime}^\ast \rangle
= \tfrac{1}{2}\delta_{n n^\prime}
, \\
\langle \delta R_n^{z}\,\delta R_{n'}^{z}\rangle
&=
\frac{1}{N_n}\,\delta_{n n^\prime}.
\end{aligned}
\end{equation}

To assess the performance of the different methods, we compute the variance of the propagated optical field normalized to the coherent-state variance, which we refer to as the squeezing ratio $\textit{Sq}_\text{ratio}$. Results obtained using the TWA are compared both to those from the PPR derived in this work, as well as from the method developed by Drummond and Raymer~\cite{Drummond:1991aa}, which we denote as $\mathrm{PPR_{DR}}$.  
Instead of the Jordan--Schwinger mapping in our PPR, $\mathrm{PPR_{DR}}$ uses an alternative mapping to stochastic variables that relies on the assumption of a sufficiently large number of atoms per spatio-frequency cell (i.e., the large-$N$ limit). A further distinction between the two implementations lies in the procedure used to factorize the diffusion matrix and construct the corresponding stochastic noise terms.

Quadrature squeezing is evaluated by mixing the propagated pulse with a shape-matched local oscillator~\cite{Lee:2009aa}. Denoting the local oscillator mode by $f_{\mathrm{LO}}(\tau)$, the quadrature operator $\hat{M}$ is defined as
\begin{equation}
\hat M_{\theta} = \int d\tau  [ f_{\mathrm{LO}}(\tau)\phi(\tau,z)e^{i\theta}  + f_{\mathrm{LO}}^*(\tau)\phi^\dagger(\tau,z)e^{-i\theta} ] \, ,
\end{equation}
where the local oscillator is normalized such that \mbox{$\int |f_{\mathrm{LO}}|^2 \, d\tau = 1$. }

For the PPR, the variance must be computed in normal order, which requires reordering of the mean square: $\hat M^2_{\theta} = :\hat M^2_{\theta}: + 1$, leading to the squeezing ratio
\begin{equation}
	\label{Sq_ratio_PPR}
S_\theta^{P} = 1 + \mathrm{Var}_{+P}[\hat{M}_\theta].
\end{equation}
In contrast, the TWA employs symmetric ordering, and no reordering correction is required,   
\begin{equation}
S_\theta^{\mathrm{TWA}} = \mathrm{Var}_{\mathrm{TWA}}[\hat{M}_\theta].
\end{equation}

We consider a coherent pulse  of duration $3.66~\text{ps}$, containing approximately  $1.83 \times 10^{14}$ photons,  interacting with an ensemble of atoms arranged in a one-dimensional array.  The atomic transition is described by a Voigt spectral profile ~\cite{Roston:2005aa} centered at the transition wavelength $\lambda_0 = 7.94 \times 10^{-7}~\text{m}$, with  full width at half maximum $\mathrm{FWHM}_{\text{Voigt}} = 5.27 \times 10^{8}$. The  effective atom–field coupling is  $g_{\phi} = 9.16 \times 10^{-8}~\text{m}^2/\sqrt{\text{s}}$.  In addition, the main numerical parameters used in the simulations are mentioned in {Table}~\ref{tab:numerical_parameters}.

\begin{figure}[t]
	\includegraphics[width=\columnwidth]{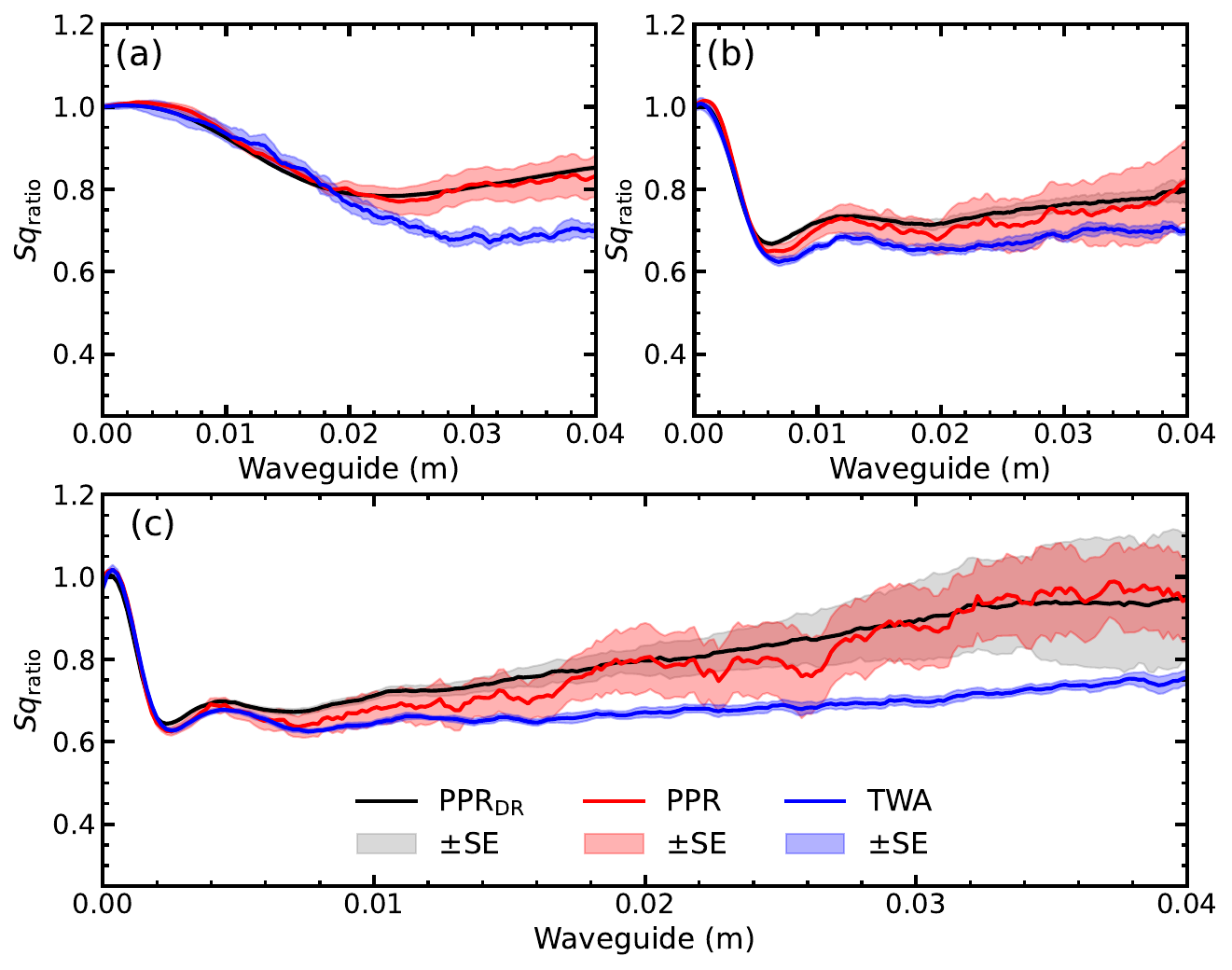}
	\caption{Comparison of three methods, i.e, PPR, TWA, and $\mathrm{PPR}_{\text{D}}$.
		The figure shows the evolution of the squeezed field quadratures within the waveguide in presence  of optical reservoir; $\kappa=10$ m$^{-1}$ for three different atom density: (a) $\rho = 2.65\times10^{21}\,\mathrm{m^{-3}}$,
		(b) $\rho = 1.33\times10^{22}\,\mathrm{m^{-3}}$, and 	(c) $\rho = 3.7\times10^{22}\,\mathrm{m^{-3}}$. The number of stochastic trajectories used is approximately {$3\times10^{5}$ for PPR, $5\times10^{4}$} for $\mathrm{PPR}_{\mathrm{DR}}$,
		and $5\times10^{3}$ for TWA.  }
	\label{fig:Fig_all_methods_Opt}
\end{figure}

\begin{table}[b]
    \centering
    \renewcommand{\arraystretch}{1.3}
    \caption{Main numerical parameters used in the simulations.}
    \label{tab:numerical_parameters}
    \begin{tabular}{lc}
        \hline
        \textbf{Parameter} & \textbf{Value} \\
        \hline
        Waveguide length $L$ & $4\times10^{-2}\,\mathrm{m}$ \\
         Waveguide diameter $D$ & $60\,\mu\mathrm{m}$ \\
        Spatial step $\Delta z$ & $1.0\times10^{-5}\,\mathrm{m}$ \\
        Number of spatial grid points & $\approx 7.5\times10^{3}$ \\
        Time window $\tau_f$ & $5.0\times10^{-11}\,\mathrm{s}$ \\
        Time step $\Delta \tau$ & $5.0\times10^{-14}\,\mathrm{s}$ \\
        Number of temporal grid points & $\approx 10^{3}$ \\
        Number of stored propagation snapshots & $\approx 200$ \\
        \hline
    \end{tabular}
\end{table}

We examine three  atomic densities--  $2.65\times10^{21}$, $1.33\times10^{22}$, and $3.7\times10^{22}\,\mathrm{m^{-3}}$-- {to} compare the three methods under both unitary evolution and in the presence of an optical reservoir under a same specific scenario.  Figure~\ref{fig:Fig_all_methods_Unitray} shows that all three methods remain in good agreement during unitary evolution,  albeit with growing sampling error evident in the $\mathrm{PPR_{DR}}$ and especially the PPR.  At higher densities,  systematic deviations of the TWA become apparent. The TWA reliance on initial-state noise alone is insufficient to accurately capture the cumulative effects of interactions on quantum fluctuations over extended propagation lengths. Evidently the TWA's approach to including the quantum fluctuations only via the initial conditions is insufficient to capture the precise effects of interactions on the quantum noise, particularly for longer interaction lengths.

The differences in stochastic behavior between the PPR and the $\mathrm{PPR_{DR}}$ shown in both Fig.~\ref{fig:Fig_all_methods_Unitray} and Fig.\ref{fig:Fig_all_methods_Opt} arise from the distinct mappings of quantum operators to stochastic variables. Provided the assumptions underlying each mapping are satisfied, both approaches reproduce the same physical predictions within statistical uncertainty. 

At higher atomic densities, increased multiplicative noise leads to larger statistical uncertainties, which can make differences between PPR and $\mathrm{PPR_{DR}}$ more pronounced.

This issue of multiple equivalent stochastic representations for a given quantum process is closely related to the concept of a \textit{diffusion gauge} ~\cite{Deuar:2002aa,Drummond:2003aa,Wuster:2017aa}, which refers to the freedom in choosing how the diffusion matrix $\mathbf{D}$ is factorized into the $\mathbf{B}$ terms. For example, in the noise terms written in this paper for PPR, one could rescale the noise contributions: making the $F^{\alpha}$ terms smaller by a factor $X$, while correspondingly increasing the $F^{\beta}$ terms by the same factor. Both factorizations correspond to the same diffusion matrix and thus the same underlying quantum dynamics, but they can lead to quite different stochastic trajectories. In principle, this freedom may be exploited to improve numerical stability—for example, by shifting multiplicative noise contributions from one set of variables to another—without altering physical predictions. 

Figure~\ref{fig:Fig_all_methods_Opt} {illustrates} the predicted squeezing ratios in the presence of optical reservoir noise, assuming a {negligible} mean thermal occupation ($\bar{n}\approx 0$)  and an   attenuation coefficient of $\kappa = 10 ~\mathrm{m}^{-1}$. {
As expected, coupling to the reservoir degrades the squeezing performance across all three methods compared to the unitary limit. Notably, the deviation in the Truncated Wigner Approximation (TWA) is particularly pronounced, underscoring its failure to accurately capture the dynamics when dissipation and nonlinear interactions simultaneously govern the evolution of quantum noise.}


\section{Concluding remarks}
\label{sec:conc}

In this work, we have employed the Jordan–Schwinger mapping to derive two complementary sets of stochastic differential equations for light–matter interactions, formulated within the truncated PPR and the TWA. These equations provide a unified framework for treating both unitary dynamics and dissipative evolution arising from coupling to an optical reservoir.

To benchmark the approximate TWA approach and the previously used $\mathrm{PPR_{DR}}$ against PPR, we studied the interaction of a coherent light pulse with a simplified two-level atomic system at resonance.  We found that, in the absence of reservoir coupling, all three methods yield quantitatively consistent results for moderate interaction strengths. As the interaction strength increases, systematic deviations emerge in the TWA, reflecting the limitations of its treatment of quantum fluctuations through initial-state noise alone.

The inclusion of an optical reservoir amplifies these differences. In this regime, discrepancies between the PPR and TWA methods become more pronounced. These findings highlight both the practical utility and the intrinsic limitations of semiclassical phase-space approaches such as the TWA when dissipation and interactions jointly shape the quantum noise dynamics.

We note that the comparison between the PPR and the TWA presented here is not restricted to one-dimensional systems. The qualitative differences between the two methods—namely the exact treatment of quantum noise in the PPR and the approximate nature of the TWA—are expected to persist in higher-dimensional settings. 

However, increasing the dimensionality leads to a larger number of coupled modes, which may enhance numerical challenges. In particular, the stability issues of the PPR can become more pronounced, while the accuracy of the TWA may depend more sensitively on the local occupation numbers and interaction strength. Therefore, while the qualitative conclusions remain valid, quantitative behavior may depend on the specific system and dimensionality.

We emphasize that the results can be considered reliable in the regime where the PPR and TWA approaches overlap and yield consistent predictions. Outside this overlap region, it becomes difficult to assess the validity of approximate methods particularly the TWA, without an independent benchmark. Had we examined only a single approach, we would lack a clear criterion for determining the parameter regime in which the results can be trusted. By comparing all three methods, however, we are able to identify the timescales for which the predictions remain robust.

In this way, the PPR provides a practical indicator of the temporal regime over which the TWA remains valid. Under the present conditions in this paper, this benchmarking allows us to establish a well-defined window in which the TWA can be trusted. In situations where such an indicator is unavailable, one should exercise caution when relying solely on the TWA, as its range of validity cannot be determined \emph{a priori} without comparison to the PPR.

The framework developed here is broadly applicable to a wide class of nonclassical radiation phenomena, including photon antibunching, SIT solitons, and quadrature squeezing during optical pulse propagation in hollow-core waveguides and related platforms. By providing a stable and flexible  PPR formulation alongside a controlled semiclassical approximation, our approach enables systematic comparisons across regimes of increasing quantum complexity.

The regime considered here is directly relevant to experiments on light propagation in dense atomic media near the resonance in photonic crystal fiber or and warm vapor cells. In particular, our results apply to near-resonant conditions where strong collective light-matter interactions lead to modified dispersion and diffraction properties.

One concrete application is the generation of nonclassical light, such as squeezing, during propagation through a resonant atomic medium. This regime has been theoretically investigated in Refs.~\cite{Najafabadi:2025aa, Najafabadi:2024aa}. While a direct experimental realization of this specific proposal is still lacking, related experiments~\cite{Vogl:2014aa,Haupl:2022aa} on nonlinear and quantum optical effects in atomic vapors operate in comparable parameter regimes.

An important direction for future work is the incorporation of collisional damping associated with the atomic reservoir. Such processes introduce additional nonlinearities into the bosonic stochastic equations and require careful treatment to faithfully capture higher-order correlations and fluctuation statistics. Extending the present framework to include these effects will be essential for a complete description of quantum light–matter dynamics in dense or strongly interacting media.

A second research direction addresses the influence of non-Gaussian dynamics. Relying exclusively on the squeezing ratio may fail to fully capture the quantum features that the TWA omits~\cite{Corney:2015aa}. Although the TWA can generate non-Gaussian statistics, its accuracy diminishes in this regime because it fundamentally neglects the negative regions of the exact Wigner function—the hallmarks of truly non-Gaussian quantum states. It would therefore be valuable to explore higher-order statistical measures and non-Gaussian signatures within the pulse propagation model.

\section*{Acknowledgments}
We are indebted to Ray Kuang Lee and  Andrei B. Klimov. L.L.S.S. acknowledges financial support from  Spanish Agencia Estatal de Investigaci\'on (Grant No. PID2021-127781NB-I00). JFC acknowledges helpful discussions with Sepanta Moussavian

{
\section*{Appendix: Derivation of the continuum propagation equation}

In this appendix, we derive the continuum form of the discrete propagation operator {appearing in Eq.~\eqref{eq:svea_paraxial}.}

We start from the discrete expression. Assuming translational invariance of the coupling kernel, {\mbox{$\Delta\omega(j,j') = \Delta\omega(j-j')$}, the operator can be diagonalized using a discrete Fourier transform. Writing $\alpha_j = \sum_k \alpha_k e^{ikx_j}$, we obtain}
\begin{equation}
    \sum_{j'} \Delta \omega (j, j') \alpha_{j'} = 
    \sum_{j'} \Delta \omega(j-j') \sum_k \alpha_k e^{ikx_{j'}}.
\end{equation}

Introducing the change of variables \(m = j - j'\), so that \(x_{j'} = x_j - x_m\), we find
\begin{equation}\label{Eq:FFT_omega}
    \sum_{j'} \Delta \omega(j-j') \alpha_{j'} = 
    \sum_k \alpha_k e^{ikx_j} 
    \sum_{m} \Delta \omega (m) e^{-ikx_m}.
\end{equation}

We define the Fourier transform of the kernel as
\[
\Delta \omega (k) = \sum_m \Delta \omega(m) e^{-ikx_m}.
\]

Since the wavevector can be decomposed as \(\mathbf{k} = (k_z, \mathbf{k}_\perp)\), we assume that the optical field is narrowly distributed around a carrier wavevector \(k_0 \hat{\mathbf{z}}\). In the slowly-varying envelope approximation, we write {$
k_z \rightarrow k_0 - i \partial_{z}$.}

Within the paraxial approximation, the dispersion relation becomes
\begin{equation}
|\mathbf{k}| = \sqrt{k_z^2 + k_\perp^2} \approx k_0 - i\frac{\partial}{\partial z} + \frac{k_\perp^2}{2k_0}.
\end{equation}

Expanding \(\omega(k)\) to first order around \(k_0\), we obtain
\begin{equation}
    \omega(k) \approx \omega(k_0) + v_g \left(-i \frac{\partial}{\partial z} + \frac{k_\perp^2}{2k_0} \right),
\end{equation}
where \(v_g = \left.\frac{\partial \omega}{\partial k}\right|_{k_0}\) is the group velocity, and we used the correspondence {$k_\perp^2 \rightarrow -\nabla_\perp^2$.}

Substituting this result {into Eq.~\eqref{Eq:FFT_omega} and} multiplying both sides by \(i\), we finally obtain the continuum propagation equation
\begin{equation}
    i \sum_{j'} \Delta \omega (j, j') \alpha_{j'} 
    = v_g \frac{\partial}{\partial z} \alpha_j 
    - i \frac{v_g}{2k_0} \nabla_\perp^2 \alpha_j.
\end{equation}
 }
%
\end{document}